\documentclass{svmult}

\usepackage{graphicx}    

\def\beq{\begin{equation}}
\def\eeq{\end{equation}}
\def\nn{\nonumber}
\def\bea{\begin{eqnarray}}
\def\eea{\end{eqnarray}}
\def\ba{\begin{array}}                  
\def\ea{\end{array}}

\def\a{\alpha}

\def\e{\epsilon}

\def\k{\kappa}

\def\s{\sigma}

\begin{document}

\title*{Gauge Theory on the Fuzzy Sphere and Random Matrices}
\author{Harold Steinacker}
\institute{Institut f\"ur theoretische Physik \\
Ludwig--Maximilians--Universit\"at M\"unchen \\
Theresienstr.\ 37, D-80333 M\"unchen, Germany
\texttt{harold.steinacker@physik.uni-muenchen.de}}
%
%
\maketitle

\section{Introduction}

Gauge theories provide the best known description of the fundamental
forces in nature. At very short distances however, physics is not
known, and it seems unlikely that spacetime is 
a perfect continuum down to arbitrarily small scales. Indeed,
physicists have started to learn in recent years
how to formulate field
theory on quantized, or noncommutative spaces.

However,  most 
attempts to (second-) quantize these field theories
using modifications of the conventional, perturbative  methods
have failed up to now. 
It seems therefore worthwhile to try to develop 
new techniques for their quantization,
trying to take advantage of the peculiarities of the non-commutative case.
There is indeed one striking feature of some non-commutative gauge theories:
they can be formulated as {\em (multi-) matrix models}.

I will explain here the main ideas of  \cite{mypaper}, 
where such a matrix formulation of (pure) $U(n)$ gauge theory on the fuzzy
sphere
has been used to calculate its partition function in the commutative
limit.
This is done using matrix  techniques 
which {\em cannot} be applied in the commutative case.

\section{The model}
\label{sec:1}

Consider the matrix model for 3 hermitian $N\times N$ matrices $B_i$, 
with action 
\beq
S(B) = \frac{2}{g^2 N} Tr\Big((B_i B^i  - \frac{N^2-1}4)^2 
+ (B_i + i\varepsilon_{ijk}B^j B^k)(B^i + i\varepsilon^{irs}B_r B_s)\Big)
\label{B-action}
\eeq
where $g$ is the coupling constant and $N$
is a (large) integer.
This action describes pure gauge theory on
the fuzzy sphere $S^2_N$, cp. \cite{madore,watamura}.
It is invariant under the $U(N)$ gauge symmetry acting as
$$
B_i \to U^{-1} B_i U. 
$$
To see that this corresponds to the usual $U(1)$ local gauge
symmetry in the classical limit,
we first note that the absolute minimum (the
``vacuum'') of the action is given by
$$
B_i = \lambda_i = \pi_N(J_i)
$$
up to gauge transformation, where $\pi_N$ is the $N$-dimensional
representation
of $su(2)$ with generators $J_i$. Upon rescaling
$\lambda_i = x_i\; \sqrt{\frac{N^2-1}{4}}$,
on finds the generators $x_i$ of the fuzzy sphere \cite{madore} which
satisfy
$$
\sum x_i x_i = 1, \quad [x_i, x_j] = i{\sqrt{\frac{4}{N^2-1}}}\; \e_{ijk} x_k.
$$
This means that the vacuum of this matrix model is the fuzzy sphere.
We can now write any field (``covariant coordinate'') as
\beq
B_i = \lambda_i + {\cal A}_i.
\label{B-split}
\eeq
Then
$$
B^i + i\varepsilon^{ikl}B_k B_l = \frac{1}{2}\varepsilon^{ikl} F_{kl}, \quad
F_{kl}:= i [\lambda_k,{\cal A}_l] - i [\lambda_l,{\cal A}_k] 
  + i [{\cal A}_k,{\cal A}_l] + \varepsilon_{klm} {\cal A}^m.
\label{F-def}
$$
Notice that the kinetic terms in the field strength $F_{kl}$ arise 
automatically due to the shift (\ref{B-split}).
The $U(N)$ gauge symmetry acts on ${\cal A}_i$ as
\beq
{\cal A}_i \to  U^{-1} {\cal A}_i U +  U^{-1} [\lambda_i, U]
\eeq
which for $U = \exp(i h(x))$ and $N \to \infty$ 
becomes the usual (abelian!) gauge
transformation for a gauge field.
One can furthermore show that the ``radial'' field 
$$
\varphi := \lambda^i {\cal A}_i
$$
decouples in the large $N$ limit, and $\frac 1N Tr \to \int$. 
Hence the model reduces to the
usual $U(1)$ Yang-Mills action
\beq
S  = \frac{1}{g^2}\int F_{mn} F^{mn}
\eeq
in the commutative (=large $N$) limit.

\paragraph{Monopoles.}
One can find explicitly new, non-trivial solutions of this model
using the ansatz \cite{mypaper} 
\beq
B_i = \a_m \lambda^{(M)}_i
\label{monopoles}
\eeq
for suitable normalization constant $\a_m$. Here 
$\lambda^{(M)}_i = \pi_M(J_i)$ is the
generator of the $M$-dimensional irrep of $su(2)$, which can
be embedded in the configuration space of $N\times N$ matrices 
if $m = N-M >0$. 
It turns out that (\ref{monopoles}) describes monopole
solutions with monopole charge $m$, and the corresponding gauge
potential can be calculated explicitly \cite{mypaper}. 
Notice that negative monopole charges $m<0$ can also be obtained 
by admitting matrices $B_i$ of size $N'\times N'$ with $N < N' \ll
2N$, while keeping the action (\ref{B-action}) as it is.

Finally, it should be noted that the correct action 
$S = \frac{m^2}{2g^2}$ of the above
monopoles is recovered only upon a slight modification of the action  
(\ref{B-action}) as indicated in \cite{mypaper}, which does not affect
the classical limit. The reason is that the 
``empty'' blocks in (\ref{monopoles}) if embedded in $N \times N$
matrices give a large contribution due to the first term in (\ref{B-action}).
This is certainly unphysical (it could be interpreted as action of a
Dirac string), and can be avoided by the slightly modified action (78)
in \cite{mypaper}. Then the energy of all monopoles is correctly
reproduced in the commutative limit $N \to \infty$. All this extends
immediately to the non-abelian case:

\paragraph{Non-abelian case}
This model is readily extended to the nonabelian case by
using matrices of size $n N$, i.e. 
$B_i = B_{i,\a} t^\a = \lambda_i\; t^0  + {\cal A}_{i,0}\; t^0 +
{\cal A}_{i,a}\; t^a$ 
where $t^a$ denote the Gell-Mann matrices of $su(n)$. The action then
reduces  to the
usual $U(n)$ Yang-Mills action
\beq
S  = \frac{1}{g^2}\int (F_{mn,0} F^{mn,0} + F_{mn,a} F^{mn,a})
\eeq
in the commutative limit. Again, all ``instanton'' sectors
are recovered if one admits matrices of arbitrary size $M \approx nN$ for the
above action.

\section{Quantization}

The quantization of $U(n)$ Yang-Mills
gauge theory on the usual 2-sphere is well known, see e.g.
\cite{migdal,witten}.
In particular, the partition function and  correlation functions
of Wilson loops have been calculated.
Our goal is to calculate the partition function
for the YM action (\ref{B-action})
on the fuzzy sphere, taking
advantage of the formulation as matrix model. This can be
achieved by collecting the 3 matrices $B_i$ into a single $2M
\times 2M$ matrix
\beq \label{fluc-C-nonabel-1}
C = C_0 + B_i\sigma^i 
\eeq
The main observation is that the above action (\ref{B-action})
can be rewritten simply as
$$
S(B) = Tr V(C)
$$
imposing the constraint $C_0 = \frac 12$, for the potential
$$
V(C) = \frac{1}{g^2 N} (C^2 -(\frac N2)^2)^2
$$
Then we proceed as
\bea
Z &=& \int d B_i\; \exp(-S(B))) \nn\\
 &=& \int d C\;  \delta(C_0 - \frac 12)\; \exp(-Tr V(C)) \nn\\
 &=& \int d\Lambda_i \Delta^2(\Lambda_i) \exp(-Tr V(\Lambda)) 
     \int dU \delta((U^{-1}\Lambda U)_0 - \frac 12) \nn
\eea
where $dU$ is the integral over $2M\times 2M$ unitary matrices,
$C = U^{-1}\Lambda U$, and $\Delta(\Lambda_i)$ is the Vandermonde-determinant of
the eigenvalues $\Lambda_i$.
Here $\delta(C_0 - \frac 12)$ is a product over $M^2$
delta functions, which 
can be calculated by introducing
$J = \left(\begin{array}{cc} K & 0 \\ 0 &  K \end{array}\right)
 = K\; \s^0$
where $K$ is a $N \times N$ matrix. Then
$$
\delta((U^{-1}C U)_0 - \frac 12) = \int dK 
\exp(i Tr(U^{-1} (C -\frac 12) U J)).
$$
By  gauge invariance, the r.h.s. depends only on the 
eigenvalues $\Lambda_i$ of $C$. 
Hence
\bea
Z  &=& \int dK  \int d\Lambda_i \Delta^2(\Lambda_i) \exp(-Tr V(\Lambda)) 
    \int dU \exp(i Tr(U^{-1}\Lambda U J - \frac 12 J)) \nn\\
 &=&   \int dK\;  Z[J]\; e^{-\frac i2 Tr J}
\label{Z-J}
\eea
where
\beq
Z[J] := \int d C\;  \exp(-Tr V(C) + i Tr (C J))
\eeq
depends only on the eigenvalues $J_i$ of $J$.
Diagonalizing $K = V^{-1} k V$, we get 
$$
Z  = \int d k_i  \Delta^2(k) \int d\Lambda_i \Delta^2(\Lambda_i) \exp(-Tr V(\Lambda)) 
    \int dU \exp(i Tr(U^{-1}(\Lambda-\frac 12) U J)) 
$$
where $\int dV$ was absorbed in $\int dU$. 
The integral over $\int dU$
can now be done using the Itzykson-Zuber-Harish-Chandra formula 
\cite{IZ2}
$$
\int dU \exp(i Tr(U^{-1} C U J)) = const 
  \frac{\det(e^{i \Lambda_i J_j})}{\Delta(\Lambda_i) \Delta(J_j)},
$$
which also depends only on the eigenvalues of $J$ and $C$.

In this step the number of integrals is reduced from $N^2$ to 
$2N$. This basically means that the integral over fields on 
$S^2_N$ is reduced to the integral over functions in one variable. 
This is a huge step, just like in the usual matrix models. The constraint
however forces us to evaluate in addition the integral over $k_i$, 
which is quite complicated due to the rapid oscillations in 
$\det(e^{i \Lambda_i J_j})$; note that $\Lambda_i \approx \pm \frac N2$. 
Nevertheless,  the integrals 
can be evaluated for large $N$ \cite{mypaper}, with the result 
$$
Z_{m} =  \sum_{m_1 + ... + m_n = m}\; 
     \int_{-\infty}^{\infty} d\k_1... d \k_n\;\Delta^2(\k) \;  e^{i \k_i m_i}\;
    \exp(-\frac{g^2}{2}\sum \k_i^2).
$$
Here we consider matrices of size $M = nN-m$, which
corresponds to the monopole sectors with total $U(1)$ charge 
$m = m_1 + ... + m_n$. 
This can be rewritten in the ``localized'' form as a weighted sum of
saddle-point contributions, as advocated by Witten \cite{witten}:
$$
Z_{m} = 
\sum_{m_1 + ... + m_n = m}\;  P(m_i,g) \; \exp(-\frac{1}{2g^2}\sum m_i^2)
$$
where $P(m_i,g)$ is a totally symmetric polynomial in the $m_i$ which
can be given explicitly.
In order to include all monopole configurations, 
we should simply sum over matrices of different sizes $M = nN -m$,
for the same action given by $V(C)$. 
One can indeed find corresponding saddle-points of the action 
(\ref{B-action})
which have the form \cite{mypaper}
$$
A_i = \left(\begin{array}{cccc}m_1 A_i & 0 & ... & 0 \\
                               0 & m_2 A_i & ... & 0 \\
                               \vdots & \vdots & \ddots & \vdots \\
                               0 & 0 & ... & m_n A_i \end{array}\right)
$$
where 
$$
\vec A_i = \vec r \times \vec {\cal A} \approx \frac m2 \frac 1{1+x_3}\;
\left(\begin{array}{c} x_2\\-x_1\\0\end{array}\right)
$$
becomes the usual monopole field for large $N$, and action becomes
$$
S(C^{(m_1, ..., m_n)}) = \frac1{2g^2}\; \sum_i m_i^2
$$
for large $N$. This is a standard result on the classical sphere.

 Hence the full partition function is obtained by
summing\footnote{the relative weights of $Z_m$ for different $m$ is 
not determined here. However, it could be calculated in principle}
over all $Z_m$, 
$$
Z = \sum_m Z_m = \sum_{m_1, ...,m_n = -\infty}^{\infty} \; 
     \int d \k_i\;\Delta^2(\k) \;   e^{i \k_i m_i}\;
    \exp(-\frac{g^2}{2}\sum \k_i^2).
\label{Z-full}
$$
Using a Poisson resummation, 
this can be rewritten in the form
$$
Z = \sum_{p_1, ..., p_n \in Z}\; \Delta^2(p) \; 
    \exp(-2\pi^2 g^2 \sum_i p_i^2)
$$
or equivalently
$$
Z = \sum_{R}\; (d_R)^2 \; 
    \exp(-4\pi^2 g^2 C_{2R}).
$$
Here the sum is over all representations of $U(n)$, $d_R$ is the
dimension of the representation and $C_{2R}$ the quadratic casimir. 
This form was found in \cite{migdal} for the
partition function of a $U(n)$ Yang-Mills
theory on the ordinary 2-sphere.

We see that the limit $N \to \infty$ of the partition function
for $U(n)$ YM on the fuzzy sphere is well-defined, and 
reproduces the result for YM on the classical sphere. 
This strongly suggests that the same holds for the full 
YM theory on the fuzzy sphere, and that
there is nothing like UV/IR mixing for pure gauge theory on $S^2_N$.
This is unlike the case of a scalar field,
which exhibits a ``non-commutative anomaly'' \cite{fuzzyloop} 
related to UV/IR  mixing.


\begin{thebibliography}{99.}
%
%
%
\bibitem{mypaper} H. Steinacker,  
\textit{Quantized Gauge Theory on the Fuzzy Sphere as Random Matrix
  Model}, Nucl. Phys. {\bf B679}, vol 1-2 (2004) 66-98

\bibitem{migdal}  A.A. Migdal, 
\textit{Recursion equations in gauge theories}, 
 Sov.Phys.JETP {\bf 42} (1976) 413;
 B.E. Rusakov, 
\textit{Loop averages and partition functions in $U(N)$ gauge theory on
two-dimensional manifolds},
Mod.Phys.Lett. {\bf A5} (1990) 693


\bibitem{madore} J. Madore, \textit{The Fuzzy Sphere}, 
   Class. Quant. Grav. {\bf 9}, 69 (1992) 

\bibitem{watamura} U. Carow-Watamura, S. Watamura, 
 \textit{Noncommutative Geometry and Gauge Theory on Fuzzy Sphere},
 Commun.Math.Phys. {\bf 212} (2000) 395  

\bibitem{witten} E. Witten, \textit{Two-dimensional gauge theories
  revisited}, J. Geom.  Phys. {\bf 9} (1992) 303.

\bibitem{IZ2} C. Itzykson, J.B. Zuber 
\textit{The planar approximation. 2},
Journ. Math. Phys.  {\bf 21} (1980), 411


\bibitem{fuzzyloop}
C.~S.~Chu, J.~Madore and H.~Steinacker,
\textit{Scaling limits of the fuzzy sphere at one loop},
JHEP {\bf 0108} (2001) 038



\end{thebibliography}
\end{document}